\documentclass[conference]{IEEEtran}

\makeatletter

\def\ps@headings{%

\def\@oddhead{\mbox{}\scriptsize\rightmark \hfil \thepage}%

\def\@evenhead{\scriptsize\thepage \hfil \leftmark\mbox{}}%

\def\@oddfoot{}%

\def\@evenfoot{}}

\makeatother

\ifCLASSINFOpdf
\else
\fi
\usepackage{graphicx}
\usepackage{url}
\usepackage{float}
\usepackage{multirow}
\usepackage{amsmath}
\usepackage{subfigure}
\usepackage{balance}
\restylefloat{table}
\newcommand{\nop}[1]{}

\begin{document}
%

\title{The Best Answers? Think Twice: Online Detection of Commercial Campaigns in the CQA Forums}

%
%
%
%
%

\author{\IEEEauthorblockN{Cheng Chen}
\IEEEauthorblockA{Dept. of Computer Science\\
University of Victoria\\
Victoria, BC, Canada}

\and
\IEEEauthorblockN{Kui Wu}
\IEEEauthorblockA{Dept. of Computer Science\\
University of Victoria\\
Victoria, BC, Canada}

\and
\IEEEauthorblockN{Venkatesh Srinivasan}
\IEEEauthorblockA{Dept. of Computer Science\\
University of Victoria\\
Victoria, BC, Canada}

\and
\IEEEauthorblockN{Kesav Bharadwaj R}
\IEEEauthorblockA{Dept. of Computer Science\\
Bits-Pilani \\
Pilani, India}

}

\nop{
\alignauthor
Cheng Chen\\
       \affaddr{Dept. of Computer Science}\\
       \affaddr{University of Victoria}\\
       \affaddr{Victoria, BC, Canada}\\
       \email{cchenv@uvic.ca}
\alignauthor
Kui Wu\\
       \affaddr{Dept. of Computer Science}\\
       \affaddr{University of Victoria}\\
       \affaddr{Victoria, BC, Canada}\\
       \email{wkui@ieee.org}
\alignauthor 
Venkatesh Srinivasan\\
       \affaddr{Dept. of Computer Science}\\
       \affaddr{University of Victoria}\\
       \affaddr{Victoria, BC, Canada}\\
       \email{venkat@cs.uvic.ca}
\and  
\alignauthor Kesav Bharadwaj R\\
       \affaddr{Dept. of Computer Science}\\
       \affaddr{Bits-Pilani}\\
       \affaddr{Pilani, India}\\
       \email{rkesav1990@gmail.com}
}

\maketitle
\begin{abstract}
In an emerging trend, more and more Internet users search for information from Community Question and Answer (CQA) websites, as interactive communication in such websites provides users with a rare feeling of trust. More often than not, end users look for instant help when they browse the CQA websites for the best answers. Hence, it is imperative that they should be warned of any potential commercial campaigns hidden behind the answers. Existing research focuses more on the quality of answers and does not meet the above need. Textual similarities between questions and answers are widely used in previous research. However, this feature will no longer be effective when facing commercial paid posters. More context information, such as writing templates and a user's reputation track need to be combined together to form a new model to detect the potential campaign answers. In this paper, we develop a system that automatically analyzes the hidden patterns of commercial spam and raises alarms instantaneously to end users whenever a potential commercial campaign is detected. Our detection method integrates semantic analysis and posters' track records and utilizes the special features of CQA websites largely different from those in other types of forums such as microblogs or news reports. Our system is adaptive and accommodates new evidence uncovered by the detection algorithms over time. Validated with real-world trace data from a popular Chinese CQA website over a period of three months, our system shows great potential towards adaptive online detection of CQA spams.             
\end{abstract}

\nop{
\begin{IEEEkeywords} 
CQA forums; Online detection; Paid posters
\end{IEEEkeywords}

\terms{Design, Experimentation, Measurement}

\keywords{CQA forums; Online detection; Paid posters} 
}

\section{Introduction}
\label{sec:intro}


Web 2.0 social websites are playing an increasingly important role on the Internet by utilizing the wisdom of crowds. One such example is the Community Question and Answer (CQA) portals on which users can post and answer questions, such as Yahoo! Answers, Naver and Baidu Zhidao~\cite{yahooanswers,naver,baiduzhidao}. Some CQA websites like Quora~\cite{quora} attract users by offering professional answers, most of which come from verified people in reality. These websites gain popularity and trust by providing a sense of interaction between the questioner and the masses. With millions of archived Q\&A sessions, CQA forums have become a major source of advice for many Internet users.    

As a large knowledge base of crowds, the archived Q\&A sessions have been used for automatic question answering and recommendation. Nevertheless, the quality of user-generated content in the Q\&A sessions varies drastically. For instance, some answers do not match the questions and even contain spam and rude words. In recent years, tremendous efforts have been made to locate better answers and remove spam from the archived questions and answers resource. Techniques such as analysis of text, user-question-answer's link relationship, and user feedback features have been used in tools like PageRank to identify high-quality web pages~\cite{Jeon:2006:FPQ:1148170.1148212,Jurczyk:2007:DAQ:1321440.1321575,Agichtein:2008:FHC:1341531.1341557}. 

Existing techniques, however, may not work well in the presence of the so-called Internet water army, a large crowd of hidden posters who get paid to generate artificial content in the social media for commercial profits. Paid posters have become popular with the booming of crowd-sourcing marketing. As confirmed in~\cite{Wang:2012:WWW}, crowd-sourcing systems such as Amazon's Mechanical Turk, Zhu Ba Jie (a similar Chinese crowd-sourcing site), have been broadly used for commercial campaigns. Due to their popularity, the CQA forums have become the targets of those campaigns that create untruthful Q\&A sessions for commercial purpose. Consider the following example:    
%

\textit{Question: I tried several methods to lose weight but all failed. What should I do? Please give me some advice!}

\textit{Best answer: Don't worry, I have experienced the same pain as you. Firstly, you have to keep a healthy diet. Be careful about the nutrition in your food and never eat  fast food. Secondly, don't sit too long in front of a computer. Finally, perform physical exercise everyday. What's more, you can also try a product named X. This product cotains ingredients such as ... and can help you lose weight without any risks.}

The above Q\&A session was actually generated by paid posters. The answer provides very practical advice at first and then gives suggestion on the product which needs to be promoted. The practical advice part is to earn the trust of the users. We have observed that fake answers generated by paid posters are often long enough and quite relevant to the questions, and some paid posters involved in the fake Q\&A sessions are ranked high according to the website's reputation system.

Based on textual similarities, previous work~\cite{Liu:2008:USA:1599081.1599144,Bian:2008:citeulike:2803519,Bian:2009:citeulike:4375490} is likely to treat the above answer as of high quality due to the high relevance of textual features between the answer and question content. As a result, the output may contain commercial spam, resulting in a credibility problem. Therefore, additional strategies, such as \emph{writing templates}, \emph{public calls for commercial campaigns}, and \emph{a poster's track reputation}, should be integrated for the effective detection of paid posters. Furthermore, most existing work relies on offline analysis, while end users demand for instant help and should be warned of potential commercial campaigns when they browse a CQA forum. The call for a real-time response system that can detect potentially fake Q\&A sessions on the fly is strong. 

We tackle the above two challenges in this paper by designing an adaptive online detection system tailored specifically for CQA forums. Our contributions are as follows: 
\begin{itemize}
	\item We discover that the behavioral features of paid posters are different in CQA forums when compared to other types of forums such as microblog (also called Weibo, a Twitter like service in China) and news reports. We identify the special features of paid posters in CQA forums that are useful in the detection.  
	\item Based on the identified special features, we design a detection method which uses machine-learning techniques and assigns credibility scores to each of the best answers by using semantic analysis and user features, such as users' history data. 
	\item We implement an adaptive, online detection system which automatically analyzes the hidden patterns of commercial spams and raises alarms instantaneously to end users whenever a potential commercial campaign is detected. Our system is adaptive and accommodates new evidence gathered by the detection algorithms over time.
\end{itemize}

\nop{NOTE:
Campaign-like best answers lead to credibility issue.
}

\section{Data Collection and Labeling} 
\label{sec:dc}

\subsection{Data Collection}

\nop{
Before stepping into analysis of spam features, we firstly introduce how we collect and label the Q\&A sessions from the CQA site, Baidu Zhidao. 
}

Since the readers tend to pay more attention to the best answers and also due to the manner in which online paid posters are supposed to work, we only collected the best answers and ignored other ones. This is to avoid collecting a large amount of irrelevant information for this study. 

In order to collect campaign Q\&A sessions, we first visited the crowd-sourcing websites, where the paid posters apply for campaign tasks and get paid, as stated in Section~\ref{sec:intro}. From the campaigns calling for paid posters, we selected $11$ closed requests because the paid posters who worked for the $11$ products had finished the tasks. We extracted keywords for the $11$ products and searched for Q\&A sessions with them on Baidu Zhidao. We used a crawler to visit and download the web pages associated with searching result. These sessions included not only the campaign sessions, but also normal sessions containing the keywords. After parsing all the collected web pages, we obtained a group of target users, including both paid posters and normal users, as well as the links to the users' homepages hosted by Baidu Zhidao. 

By following the users' homepages, we could find useful information for our research. For example, a user's homepage provides the Q\&A sessions where this user posted his/her answers (the question answering records). The question-answer history provides a good knowledge on the multiple campaigns that a potential paid poster might have been involved. Having obtained the initial dataset of IDs and links, we then visited each user's homepage, retrieved every Q\&A session that the user participated in. We only collected the closed Q\&A sessions (i.e., the best answer determined). A closed Q\&A session implies that users can no longer post new answers to the question, but they can click the ``Like" button to support the posted answers, including the best answer and other answers. From those Q\&A sessions, we finally extracted information used in our analysis. The recorded information from those web pages includes \textit{questioner ID}, \textit{answer ID}, \textit{time}, \textit{title}, \textit{question content}, \textit{answer content}, \textit{user feedbacks (visited times, ratings)}.

\nop{Since a paid poster could participate in multiple campaigns, we could find all the records by using the question-answering history.}

\nop{This could be fixed if we have a quite large dataset and can take advantage of information from other users' answering records. A complete list of such history records of each user is significant when applying user-interaction features analysis.} 

\nop{
The site also provides other user-specific properties, like answer accepted rate and answer recommended rate. As what will be presented in the Section~\ref{sec:sfo}, we don't use these user-specific features and we leave the problem how to integrate them into our framework as future work.
}

From the Q\&A website, \textit{Baidu Zhidao}, we crawled, $6462$ users' question-answer history records accumulated during a three-month period from October to December in 2011. For each user, we built a list of history information, showing the question, answer, participated user IDs, and other features. Associated with the $6462$ user IDs, we have $75,200$ Q\&A sessions in total, all having the best answer. 

\subsection{Manual Data Labeling}

To get a sample dataset for feature analysis, campaign sessions should be differentiated from the normal ones. By reading the best answers, we manually labeled the Q\&A sessions in the dataset. The labeling process mainly depends on the Q\&A templates from the crowd-sourcing websites such as Zhubajie~\cite{zhubajie} and Tiancaicheng~\cite{tiancaicheng}. We summarize the applied techniques below:

\nop{A concise definition of spam should be given before we present our method.} 

\nop{
Most of the research work in Section~\ref{sec:rw} assumed that the most relevant answers should be the best ones and deserved higher ranking scores. However, according to our survey, as well as \cite{Wang:2012:WWW}, new types of campaign Q\&A sessions could appear in the CQA websites. Posting questions and answers, the users who are paid by companies hype their products to improve the popularity. In order to attract normal readers, paid posters are often ordered to write rather relevant contents which look very professional. The criteria for judging campaign is always subtle. Our work will be done on a limited scope, that is detecting the advertisement spam, especially the fake experience spam. Campaign sessions usually fool people by fake experience. 
}

\begin{enumerate}
\item Since we have collected a list of $11$ products which were hyped in the Baidu Zhidao, we could compare the Q\&A content with the campaign templates. If the product's name is in the $11$ initial samples and the contents match the templates, such as the descriptive words and the organized pattern of sentences, we labeled it as a campaign Q\&A session. We stress that there is difference between our work and related research which needs to judge the quality of answers. The evaluation of quality of answers is usually based on question-answer relevance, length of the texts, grammar correctness, politeness, and so on. To obtain a reliable dataset, researchers often rely on multiple assessors and are faced with the difficulty of reaching an agreement among the multiple evaluation results. Our labeling method differs from the above and largely avoids the annotation difficulty, because we know exactly the name of the hyped product and how paid posters would write the Q\&A sessions. 

\item When we encountered new products not in the list of  $11$ initial samples, we recorded the product's name and searched it in the crowd-sourcing websites. If we found the template of this product, we use the above method to compare their contents.

\item If a new product is listed in the campaign websites but the template is not available, we followed some special features normally found in Email spam to make a decision. For example, a spam may use different fonts to write the telephone numbers and insert special characters between the product's name. This type of operations is usually used to escape detection by the filter system. We labeled the session as campaign if the product's name is in a campaign list and the best answer has special features similar to Email spam. 

\nop{\item We also notice that some paid posters ask and answer the questions by themselves. If the same user ID asks and answers the same question, we label it as campaign.}

\item If we could not find the new product in the campaign websites, we then tried to identify potential templates used in the same category of products and special features obvious in an Email spam. If none of those could be identified, we labeled the session as a normal session. 

\nop{
\item If there is a new product for which we don't have the template, then we will use templates of the same category. For example, there are two medicines, A and B but we only have the Q\&A  template for A. Since both of them are medicines, we assume the Q\&A sessions of B, if they were  campaign sessions, would have the similar structure and pattern as the templates of A. If this is the case, we label the sessions of B as campaign.
}
\end{enumerate}

Up to now, we have labeled $4998$ samples in our dataset. Among these, $2147$ samples are campaign Q\&A sessions and the other $2851$ samples are normal ones. The sample size is large enough for our current study. Since we selected 11 campaigns, which were posted on the crowdsourcing websites, as the seeds of our crawler, the proportion of campaign sessions is relatively high in the dataset.

When we manually labeled our datasets, we carefully read the contents of a user's post. The meaning can be understood by human but is hard to use in machine learning based classification. Even with the above template based labeling method, it is not easy to write an algorithm to automatically identify a campaign session because a poster may re-phrase the template in their own words. Due to these reasons, we need to search for statistical features that can be effectively used towards building a detection system.   

\nop{We don't use automatic method to check the product's keywords because we noticed that some sessions who have the keywords are not campaigns. We didn't find efficient methods that could exactly detect the difference. For example, the question asker consults about a product which is recommended by many paid posters on the site. The answerer, however, may already notice the campaign Q\&A sessions and he/she would tell the asker not to believe in those sessions.

What's more, we emphasize that our research is to evaluate the credibility of the best answers. In our work, we assume that a campaign Q\&A session has the credibility problem. On the contrary, an irrelevant answer will be assigned a high credibility scores (or a low campaign score) if the content is not commercial campaign. Consider the following examples:

\textit{Question: How to become a computer expert ?}

\textit{Best answer: Simple, just close your eyes and start dreaming.}

Although the best answer seems like joking, the credibility score will be high if it is not a commercial campaign.
}

\section{Analysis of Statistical Features}
\label{sec:sfo}
\nop{
To identify effective features which contribute to the credibility of Q\&A sessions, we discuss how we choose candidate features in this section. 
}
\subsection{Insufficiency of Existing Statistical Features}

We firstly demonstrate the difficulty of the problem we study by analyzing existing features, some of which have been used in related research such as evaluation of high quality answers or detection of Internet water army in news report websites~\cite{Chen:2011} and showing their limitations. 

\subsubsection{Interval Post Time}

In~\cite{Mukherjee:2012:SFR:2187836.2187863}, Arjun~{\em et al.} defined several spamming indicators for modelling the behaviour of fake review writers. They found that spammers of a spam group tend to post reviews during a short time interval. This feature has been shown to be a good indicator to detect Internet water army in news report websites~\cite{Chen:2011}.   

In our work, we consider two time stamps for a Q\&A session: One is the time when the questioner post the question topic (the ask time), and the other one is the time when the best answer is posted by a replier (the best answer posted time). We define \textit{interval post time} as the latter time stamp minus the former one.  

\begin{figure}[!ht]
\centering
\includegraphics[width=3in]{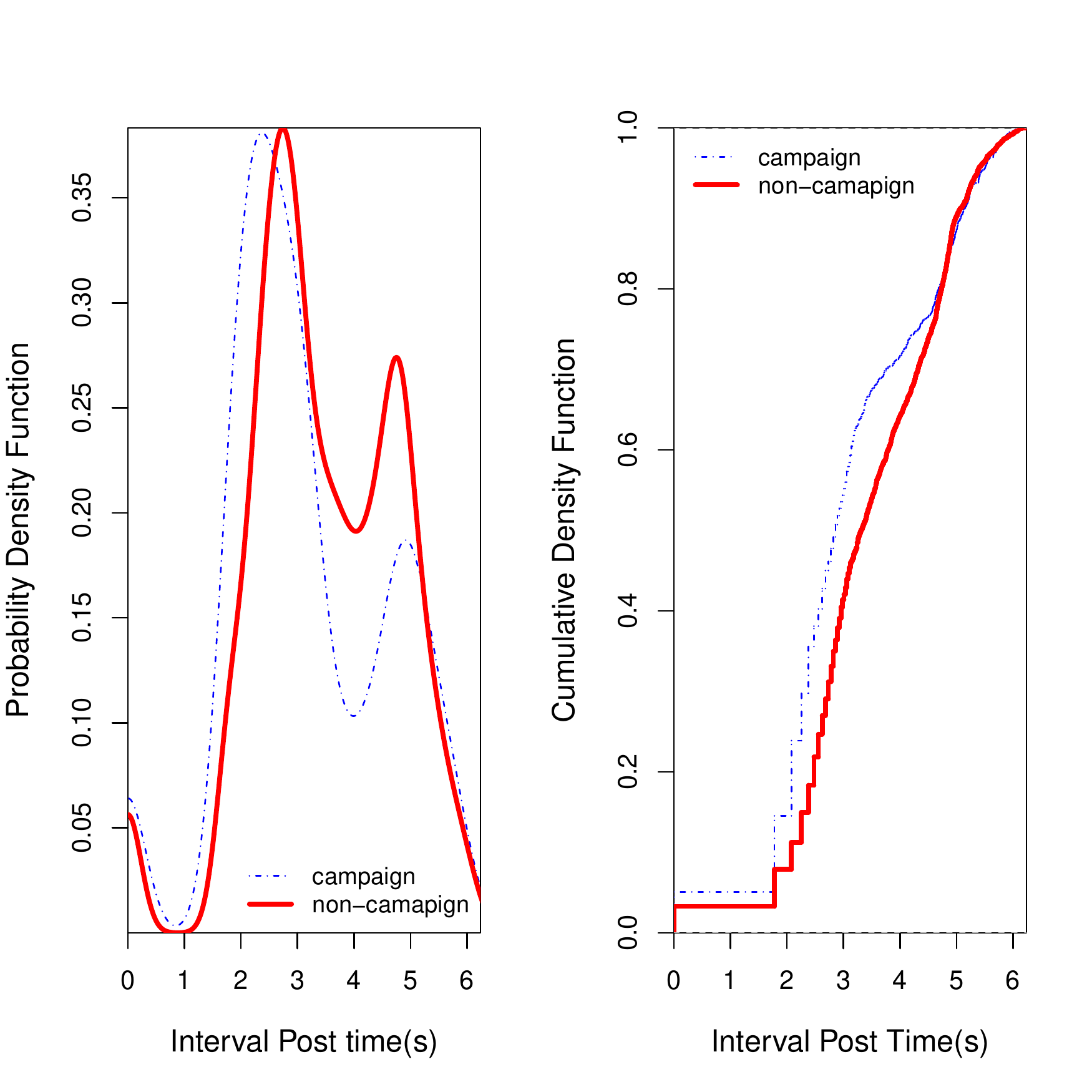}
\caption{The PDF and CDF of the interval post time}
\label{fig:it}
\end{figure}

In Figure~\ref{fig:it}, we show the approximated probability distribution of interval post time with dot-dashed lines for campaign sessions and solid lines for non-campaign sessions. The x-axis is drawn by $log$ scale.

From the figure, we find it difficult to tell the difference between campaign and non-campaign Q\&A sessions. Two reasons may contribute to the above phenomenon. There are many normal users who spend much time on the Q\&A website and try to post answers to \textit{open} questions, especially those questions associated with some \textit{rewards points}. These people are known as \textit{bounty hunters}. Most bounty hunters post very good answers because they want to get more rewards points. On the other hand, online paid posters, before they post and choose the best answer, normally wait for some random time for other answers appearing in the session. This is to give readers a fake impression that the best answer is selected among many answers. While paid posters try to finish a job as quickly as possible in news review websites~\cite{Chen:2011}, the same behaviour does not exist here.      

\nop{
The site also provides other user-specific properties, like answer accepted rate and answer recommended rate. As what will be presented in the Section~\ref{sec:sfo}, we don't use these user-specific features.
}

\subsubsection{Number of Other Answers}
\begin{figure}[!ht]
\centering
\includegraphics[width=3in]{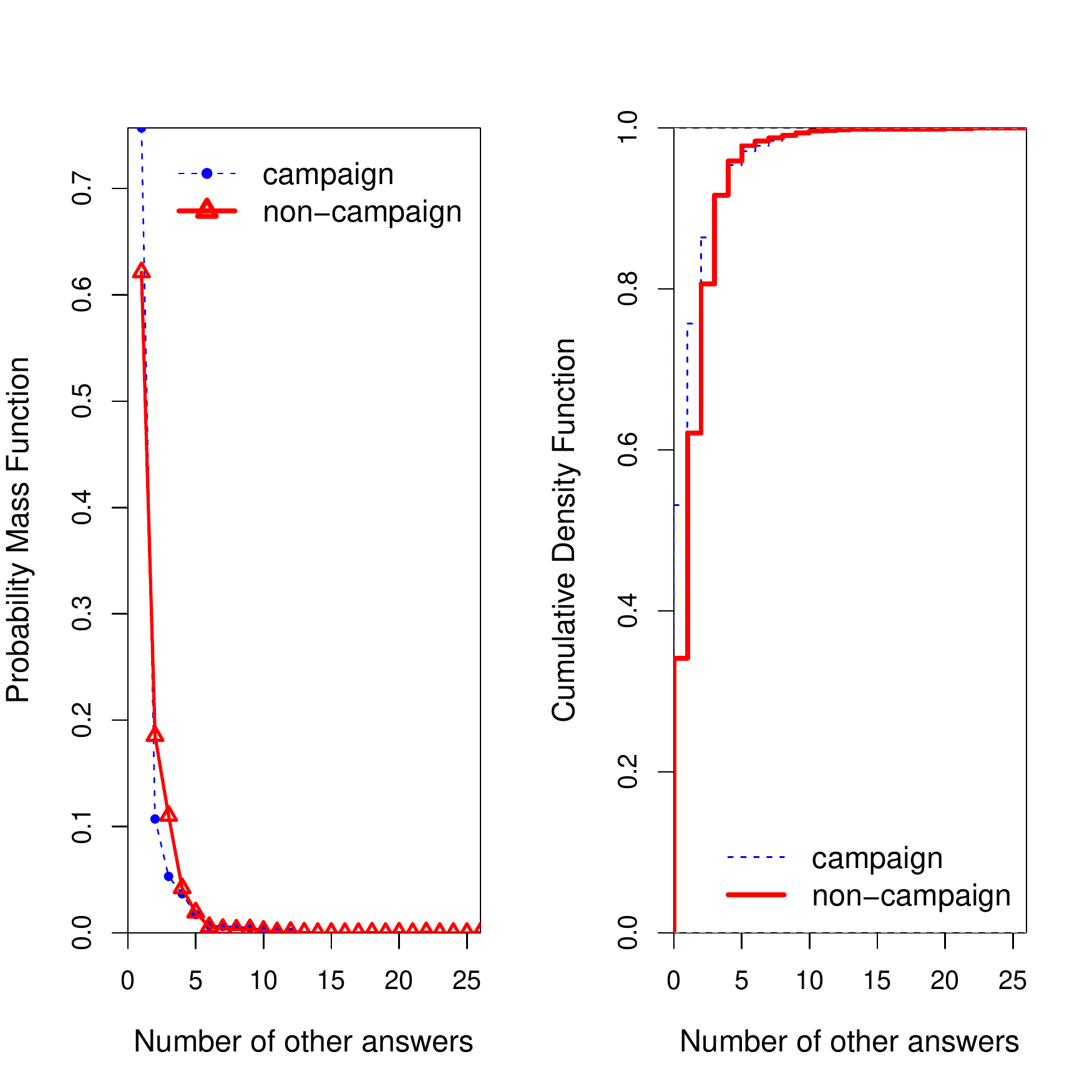}
\caption{The PMF and CDF of the number of other answers}
\label{fig:oa}
\end{figure}

Before the question is closed, users can post their own answers. This variable counts the number of answers other than the best one. Intuitively, if the paid posters create the sessions themselves, they may not have patience to wait for more replies. They could close the sessions and get paid as soon as possible. To test this conjecture, we show the probability distribution of this feature for campaign sessions and normal sessions in Figure~\ref{fig:oa} .

Similar to the interval post time, the number of other answers does not indicate much difference for the two types of Q\&A sessions. This invalidates the above conjecture and we do not consider it a good feature for the detection of paid posters in CQA portals.

\subsubsection{Number of Likes}
\begin{figure}[!ht]
\centering
\includegraphics[width=3in]{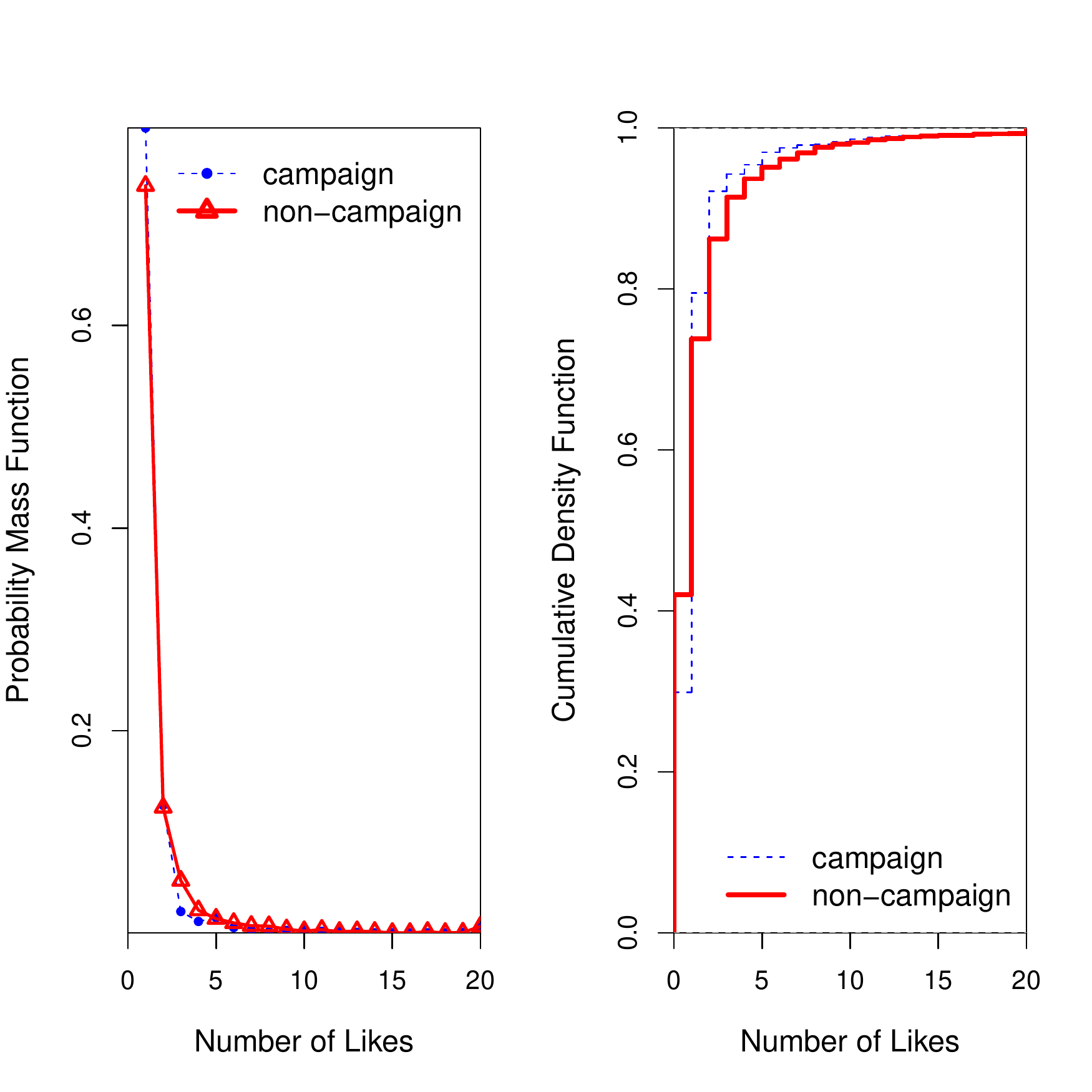}
\caption{The PMF and CDF of the number of likes}
\label{fig:likes}
\end{figure}

Similar to the ``Like" button in Facebook, if other readers find the best answer to be helpful, they may click the ``like'' button. The number on the button indicates the total number of clicks. Intuitively, this feature represents user's feedback and should be helpful in identifying trustful answers. The more ``likes" an answer receives, the more likely it is a good answer. However, as shown in Figure~\ref{fig:likes}, this is not a reliable feature. This is because the paid posters could click the button themselves and even use different user IDs to click multiple times. This behavior is also confirmed in~\cite{Bian:2008:FBV:1451983.1451997} as the ``vote spam attack".

\subsubsection{Relevance between Questions and The Best Answers}
This feature is extensively used before in identifying high-quality answers~\cite{Liu:2008:USA:1599081.1599144, Bian:2008:citeulike:2803519, Agichtein:2008:FHC:1341531.1341557, Bian:2009:citeulike:4375490}. 
The previous work is usually based on following assumptions:
\begin{enumerate}
\item Semantically high relevance between questions and answers indicates high quality.
\item Selected best answers should have higher quality than other answers. 
\end{enumerate}
The above assumptions are risky for the detection of potential campaigns created by paid posters. In commercial campaigns, answers with \textit{high-quality} are rather misleading and would beat the retrieval mechanism. Many of the answers are well-organized and highly related to the questions. In this sense, a ``high-quality'' answer does not necessarily mean trustworthiness. Thus, we do not consider the relevance measure in our work.

\subsection{Special Features for CQA Portals}
\label{sec:feature}
The limitations of existing statistical features shown above led us to look for new features specific to users in CQA websites.
\subsubsection{Spam Grade of Questioner ID (SGqID)}
It indicates whether the questioner tends to ask campaign questions. For a given questioner ID (qID), we calculate the ratio of the number of campaign sessions and the total number of sessions in which the user has participated,  
\begin{equation}
SGqID = \frac{q_1}{q_0+q_1}
\end{equation}
where $q_0$ and $q_1$ are the number of non-campaign and campaign sessions where the user appears as the questioner, respectively. To avoid $0$ probability, we specify $0.5$ to $q_1$ when $q_1 = 0$. If the system does not have enough information for a certain user(i.e., the denominator is less than $5$), we set its SGqID value to $0.5$. \footnote{This decision follows the Maximum Entropy Principle~\cite{Kapur}, i.e., we should ``make use of all the information that is given and scrupulously avoid making assumptions about information that is not available."} 

\nop{The interaction graph could be different because ours are based on a credibility foundation.}

\subsubsection{Spam Grade of Answerer ID (SGaID)}
It indicates whether the best answer poster tends to write campaign answers. For a given answerer ID (aID), we calculate the ratio of the number of campaign sessions and the total number of sessions in which the user has participated, 
\begin{equation}
SGaID = \frac{a_1}{a_0+a_1}
\end{equation}
where $a_0$ and $a_1$ are the number of non-campaign and campaign sessions the user appears as the poster of the best answers, respectively. Similar to SGqID, to avoid $0$ probability, we specify $0.5$ to $a_1$ when $a_1 = 0$. If the system does not record enough information, we set its SGaID value to $0.5$. 


\subsubsection{Spam Grade of the Text (SGtext)}
It indicates whether the collection of words in sessions associated to a user tends to be campaign specific. To calculate this feature, we need to perform statistical analysis over the words. Text information of a Q\&A session consists of the title, the content of question, and the content of the best answer. We remove the duplicate words so that we can get a collection of distinct words, $word_1$, $word_2$, $word_3$ ... $word_n$, for each Q\&A session. For each word, we calculate \textit{spam grade} which characterizes the property of the word, i.e., whether it is more campaign oriented or non-campaign oriented. Words with higher benchmark are more likely to imply hidden promotion behavior. To get rid of the impact of different length, we take the average value over the summation of the benchmarks of all words as the spam grade of the whole text. For each word, the definition of spam grade goes like this:
\begin{equation}
SGword_i = \log\left(\frac{N+1}{n_i+1}\right) * \frac{s_i+1}{S+1}
\end{equation}
where \textit{N} and \textit{S} are the total number of non-campaign and campaign sessions in the databases and \textit{$n_i$} and \textit{$s_i$} are the number of non-campaign and campaign sessions where the $word_i$ appears. The term ``\textit{$+1$}" is used to normalize the result in case of zero counts. Then the spam grade of text with $L$ distinct words is calculated as:
\begin{equation}
SGtext = \frac{SGword_1+SGword_2+ ... + SGword_L}{L}
\end{equation}

\nop{
\subsection{Other features}
Other features like user reputation, accept ratio. We don't directly use them but extract the information on our own since we can not only extract the sessions in the profile but also the semantic information will be useful to us. If we use the system-calculated scores, that would not be consistent to the records in their profile. Even though it may show some values, it cannot replace the semantic analysis which can only be done on the existing sessions on their profiles. So we just calculate the score by ourselves, if applicable and necessary. We need to find a mapping from malicious sessions to the ratio. Note that a low acceptance ratio doesn't necessarily indicate the abuse of spam. Acceptance ratio doesn't help us find well-formed spam. In other words, trustworthy is different from the quality of best answers. 

According to Jurczyk {\em et al.}~\cite{Jurczyk:2007:DAQ:1321440.1321575}, the distribution of users feedback was not even and might remain sparse for unpopular topics. They showed till Jan 2007, fewer than $35\%$ of all questions had any user votes cast for existing answers, in a sample of Yahoo! Answers.
}

\subsection{Property of the Feature Set}

\begin{figure}[!ht]
\centering
\includegraphics[width=3in]{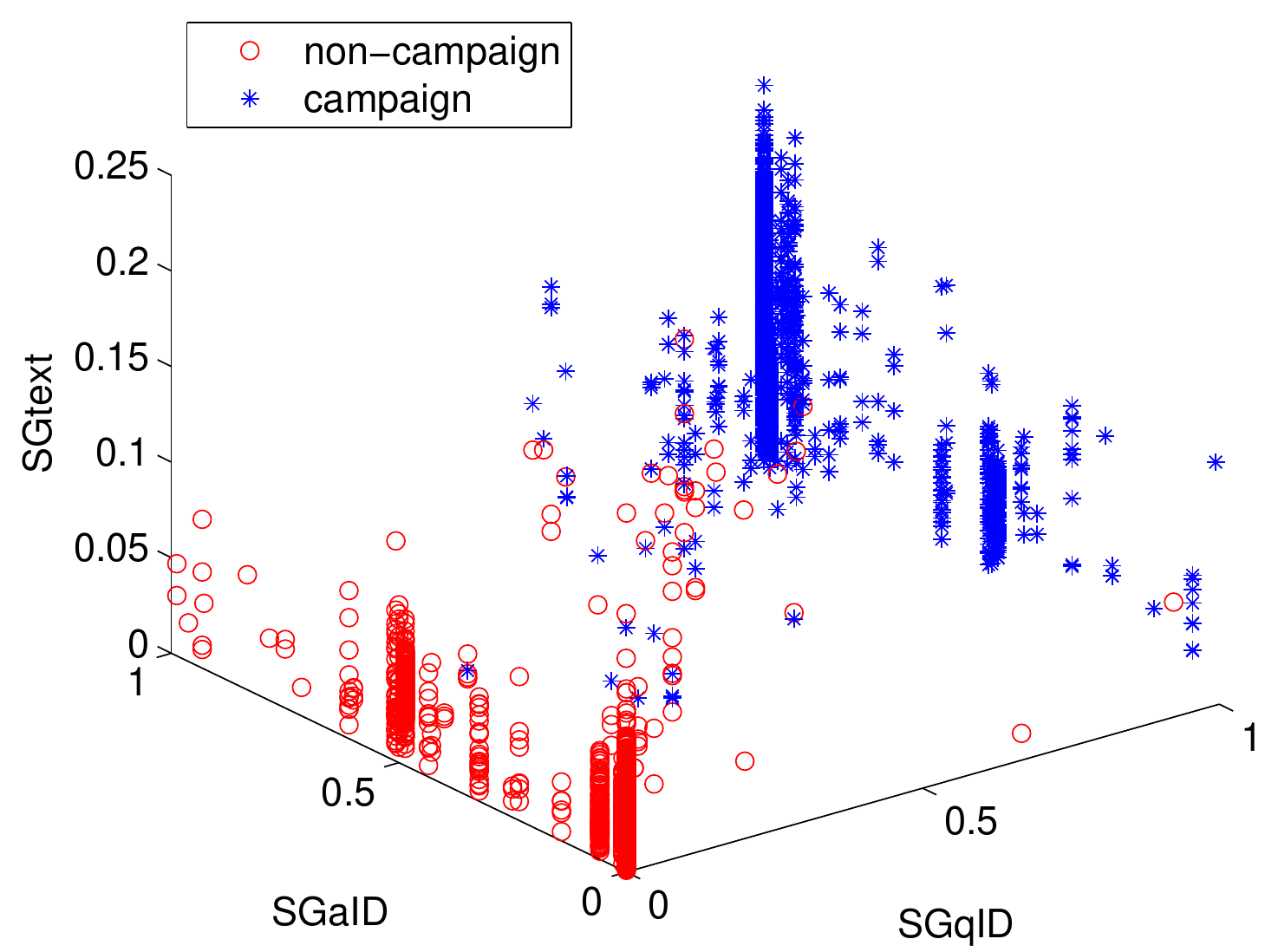}
\caption{4998 samples captured by SGqID, SGaID and SGtext}
\label{fig:4998_f}
\end{figure}

Figure~\ref{fig:4998_f} exhibits the values using the three ``SG'' features in the previous section. 

Through this figure, we can observe a clear gap between the campaign sessions and non-campaign sessions. We can then apply regression based techniques to calculate the campaign score, which indicates whether a Q\&A session tends to be a campaign.  

\section{Detection Method}
\label{sec:dsd}

In this section, we introduce a logistic regression approach to calculate campaign scores for Q\&A sessions using the three proposed ``SG" features.

\subsection{The Algorithm}
Figure~\ref{fig:4998_f} has already shown that the samples can be distinguished by the three proposed features, SGqID, SGaID and SGtext. In order to get a score indicating whether a Q\&A session is a potential commercial campaign or not, we apply logistic regression as the learning method.  We can use it to calculate values of $P(Y=1|X,\theta)$ and $P(Y=0|X,\theta)$. Here, $Y$ is a indicator variable, where $Y=1$ and $Y=0$ represent campaign and non-campaign Q\&A sessions, respectively. $\boldsymbol{X}$ is a vector of three features for each session. $\boldsymbol{\theta}$ is a vector of model parameters, each associated with a session feature and including an individually constant item(also called intercept term) which is not related to the session features.


By applying the sigmoid function, the hypothesis $h_{\boldsymbol{\theta}}(\boldsymbol{X})$ which outputs a score of $P(Y=1|X,\theta)$ or $P(Y=0|X,\theta)$ (termed as \textit{campaign score}) is defined as follows:
\begin{equation}
h_{\boldsymbol{\theta}}(\boldsymbol{X}) = \frac{1}{1+e^{-\boldsymbol{\theta^T X}}}
\end{equation}
where $\boldsymbol{\theta^T X} = \theta_1 + \theta_2*SGqID + \theta_3*SGaID + \theta_4*SGtext$. To facilitate the matrix calculation, we add an all-$1$ column to $\boldsymbol{X}$.

In practice, the higher the score, the higher the probability that the given session is a campaign session. The values of $\boldsymbol{\theta}$ will be learned by logistic regression.   
The objective then becomes an regression problem where we optimize the model so that the output campaign scores of sessions are close to their true labels ($0$ or $1$). 

The convex cost function of this optimization problem is given by
\begin{equation}
J(\boldsymbol{\theta}) = \frac{1}{m} \Sigma_{i=1}^m [-y^{(i)}log(h_{\boldsymbol{\theta}}(x^{(i)})) - (1-y^{(i)})log(1-h_{\boldsymbol{\theta}}(x^{(i)}))]
\end{equation}
where $m$ is the number of samples in the training dataset and $x$ is a matrix consisting of $m$ feature vectors of the training samples. We use gradient descent method to find the minimum of the cost function and the corresponding values in $\boldsymbol{\theta}$.


\subsection{Classification Threshold}
\begin{figure}[!ht]
\centering
\includegraphics[width=3in]{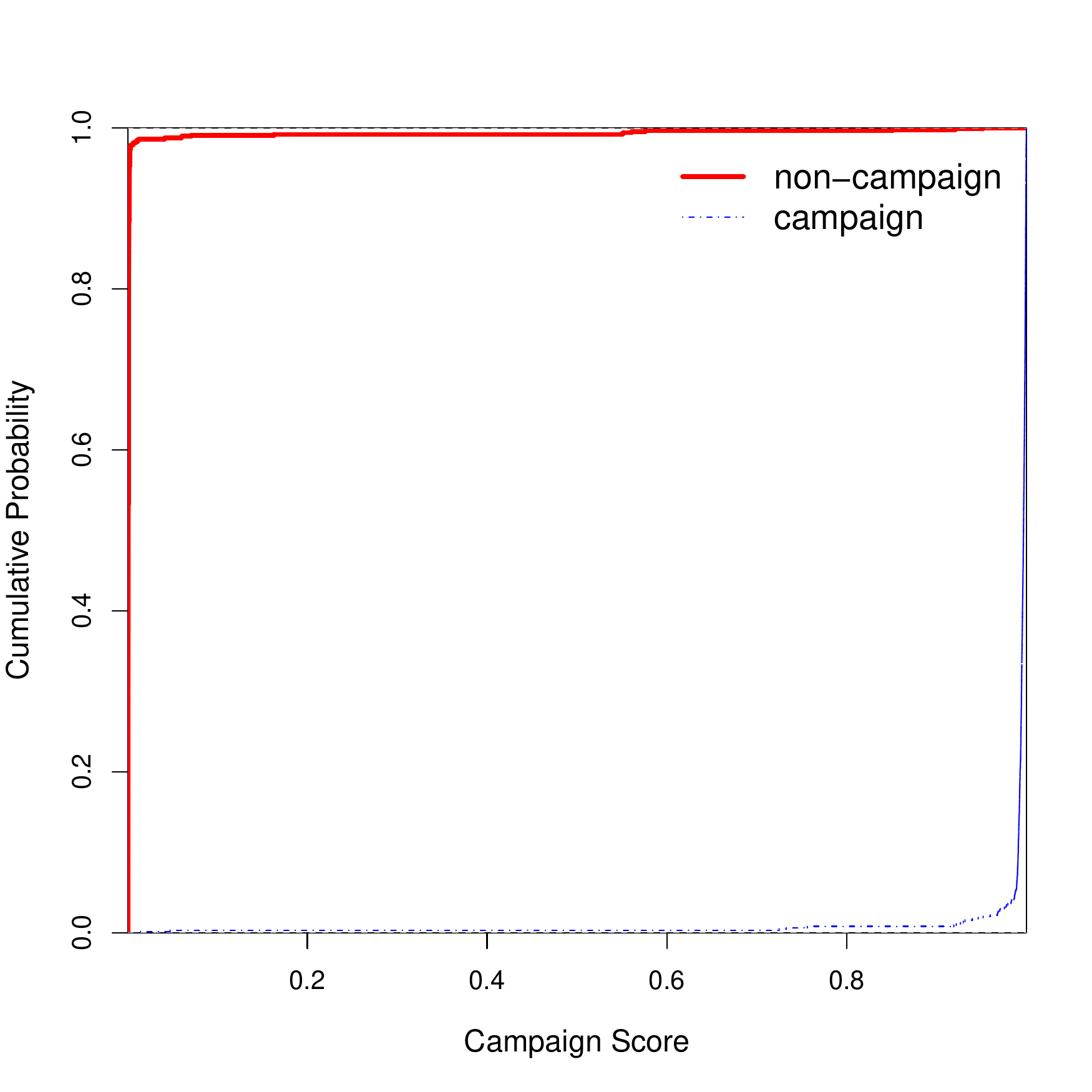}
\caption{Scores of test set (CDF)}
\label{fig:cdf_test_score}
\end{figure}

The value of $h_{\boldsymbol{\theta}}$ should be carefully determined. We shuffled the $4998$ labeled samples and took $3500$ of them as training set and the remaining $1498$  as test set. Note that the split of the dataset is arbitrary so that we can observe a suitable threshold value. When the $\boldsymbol{\theta}$ is optimized, we then calculate the campaign score of each Q\&A session in the test dataset. The distribution of scores for normal sessions and campaign sessions is shown in Figure~\ref{fig:cdf_test_score}.

\vspace{-0.1in}

From the figure, we can see that the two types of Q\&A sessions exhibit great difference on the distribution of the campaign scores. Most of the campaign scores are very close to their true labels ($Y=1$ or $Y=0$). Using the scores, we can either provide the raw scores to the users to help them make decisions when reading the answers, or we can assign the labels based on a threshold value, i.e., $Y=1$ when the campaign score is larger than the threshold value and $Y=0$ otherwise.  

\vspace{-0.1in}
\begin{figure}[!ht]
\centering
\includegraphics[width=3in]{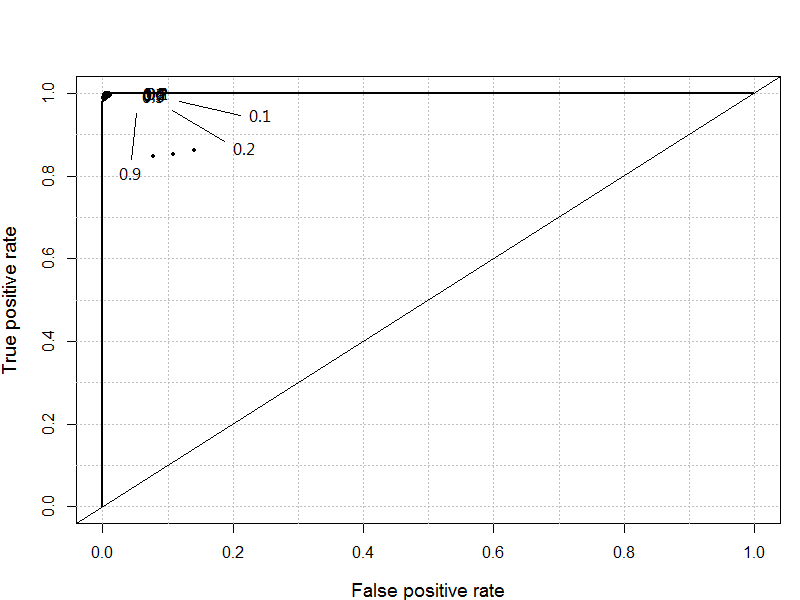}
\caption{ROC Curve of the classification result with different threshold values}
\label{fig:roc_test}
\end{figure}

In Figure~\ref{fig:roc_test}, we show the ROC curve based on different threshold values, $0.1$, $0.2$, ..., $0.9$. The points on the curve are mostly located at the top left position of the curve. The reason is that the campaign scores of most campaign sessions are higher than $0.9$ while the campaign scores of most normal sessions are smaller than $0.1$.  This curve shows that the system performance is robust with a large range of threshold value.

Based on Figure~\ref{fig:cdf_test_score} and Figure~\ref{fig:roc_test}, we set $0.5$ as our threshold for $h_{\boldsymbol{\theta}}$. We also tested it with other randomly shuffle as well as with samples sorted by timestamp and we observed similar phenomenon.

%

\section{Adaptive Online Detection System}
\label{sec:aods}

In the previous section, we have shown that we can build a model to effectively calculate the campaign score and predict the labels of unknown sessions. In practice, however, this offline analysis does not work well for users who would like to be advised of potential campaigns in real time. This requirement encourages us to design an online version of detection system, which can return campaign scores and/or predicted results in real time. We therefore build a prototype of such an adaptive online detection system. The word ``adaptive" implies that this system can update its database using new samples and generate new model parameters. 

\subsection{Overview of System Design}
\begin{figure*}[!ht]
\centering
\includegraphics[width=4.5in]{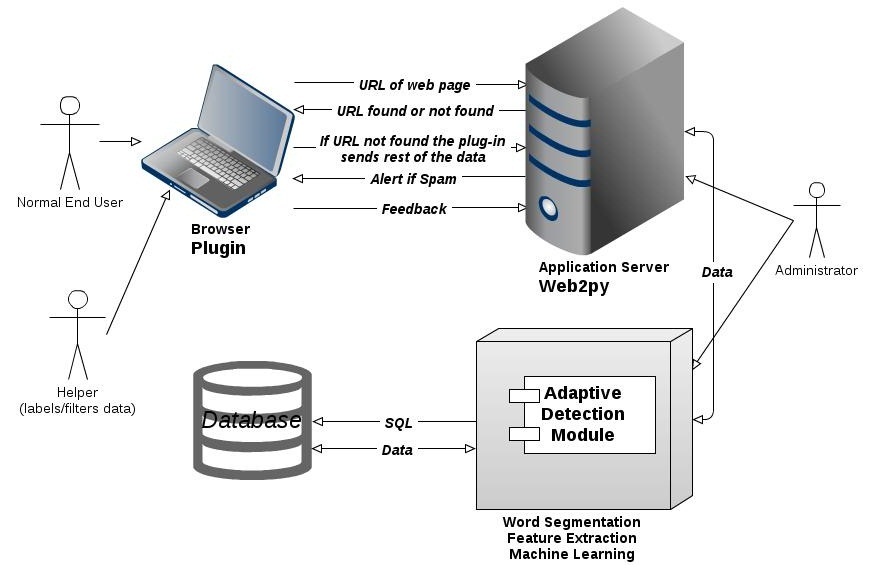}
\caption{System architecture and communication between the client and the server}
\label{fig:system}
\end{figure*}

The major components of the detection system include browser plugin and a remote server.
Figure~\ref{fig:system} shows the system architecture and the communication between the client plugin and the server. 

As shown in Figure~\ref{fig:system}, the sequence of actions that take place when a user opens a Q/A session are:
\begin{enumerate}
\item The plugin first sends only the URL of the page to the server. The
server searches for the url in its database. If it is found, the server
returns the score (spam rating) to the client. The client side script
displays the result. This avoids unnecessarily sending complete web page to the
server if it is already present in the database.
\item If the URL is not present, the server sends a response \textit{not found}
and the client after receiving the response sends the rest of the
data to the server through another \textit{XMLHTTPRequest} and waits for the
server's response.
\item The server receives the data,  segments the text into words,
and stores it in the database. The server then extracts the statistical
features necessary for the analysis from the data. Logistic regression analysis is performed to predict the class of the session (spam or no spam). If the session is classified as a spam, an alert is returned back to the user. 
\item The client-side script displays the result to the user. 
\item (Optional) If the user is an authorized user, the user can provide feedback to the server (whether or not he/she feels the session is a campaign session). There are three types of users in the system: regular users are those who use our system and they are not granted the right to annotate sessions; helper users are those who have experience and are capable of helping label the data; the administrator is the person responsible for the management of the system. 
\item When newly labelled sessions are available, the system updates the detection model using existing and newly labelled data. Note that this step could be done regularly in a  daily or even weekly basis.    
\end{enumerate}

\subsection{Plugin Design}
The plugin is a Google Chrome extension. It must be installed on the
Chrome browser in the user's system. The plugin consists of \textit{manifest.json}
file, a HTML file and a \textit{contentscript.js} file. The \textit{contentscript.js} file specifies
the javascript to be executed on the webpage the user is browsing.
The \textit{manifest.json} file contains information regarding the name, version
of plugin and the HTML, script files associated with the plugin. The manifest file also contains a list of permissions that the plugin might use to access servers. The functions of the plugin can be separated into three major steps.
\begin{enumerate}
\item Extract data from the webpage. All the data required from the webpage are extracted from the HTML
source of the webpage. Separate javascript functions were written
for extracting various information. The information extracted includes
the page URL, Question, Questioner Name, Questioner URL, Time of posting
question, Question Category, Best Answer, Answerer Name, Answerer
URL, Time of posting answer, and Rating of the answer. All the functions
are written in the \textit{contentscript.js} file.
\item Send data to the server. The server processes the data and returns
the result. The client-side Javascript communicates with the server by sending
a \textit{XMLHTTPRequest}. The \textit{POST} method is used to send the request because
the data to be sent may be big for using the \textit{GET}
method. Also for data extracted from the \textit{zhidao.baidu.com} website
the encoding of the data is set to \textit{gb2312} in order to encode
Chinese characters. 
\item The result is displayed to the user. If the user is an authorized user,  the user can enter his/her feedback to the server.
\end{enumerate}

\subsection{Server Design}
The server communicates with the plugin and also maintains a database system. The database system stores the information of Q\&A sessions and the prediction model. The server receives the Q\&A session data sent from the browser plugin. If the database has the label for the session, the server returns the label. If it is a new session, the server stores it in a buffer, calculates the spam grade based on the current model parameters, and returns the spam grade if necessary (i.e., a campaign session is detected). When enough data has been collected, we can use the \textit{helpers} to label the data. Using logistic regression, the detection model will be updated using previous data as well as the newly labelled data.

\nop{
\subsubsection{Problems}
\begin{enumerate}
\item Some of the data in the webpage were dynamic and hence were not loaded initially. Hence they were not available in the source code of the page. In order to get such data the document.readystate = ``complete" condition is checked. This makes sure that the document is completely loaded before starting to extract the data.
\item The data in the ``zhidao.baidu.com" website was of the encoding `gb-2312'. So the encoding of the XMLHTTPRequest was set to `gb-2312'. On the server side the data is first decoded using `gb-2312' and then encoded to `utf-8' because the database uses `utf-8' encoding.
\end{enumerate}
}

\subsection{Evaluation of Adaptive Online Detection System}
To evaluate the performance of adaptive online detection system, we use the collected data from \textit{Baidu Zhidao} and replay the data in multiple iterations to simulate a real-world scenario. In particular, we pretend that initially we only have partial data and use the data as the training dataset to build a detection model. In each iteration, we add some new sessions and use them as the test dataset to test the performance of the detection system. At the end of an iteration, the new sessions are added into the training dataset, and the detection model is updated using the new training dataset. This step corresponds to the scenario that new data are labeled and added into the system. Then we repeat with another iteration. Note that we sort the Q\&A sessions according to the timestamp when a session is closed. In this way, the performance is closer to that of a real-world scenario.

For the test, we begin with $200$-sample training set and build an initial detection model. At each iteration, we add $200$-sample test set. After evaluating the detection performance, we expand the training dataset with the $200$ test samples, and update the detection model with the new training dataset. We repeat this process until we use up all $4998$ samples. 
We evaluate the following four performance metrics: 
\begin{align}
Precision &= \frac{True Positive}{True Positive + False Positive}  \nonumber\\
Recall &= \frac{True Positive}{True Positive + False Negative} \nonumber\\
F-measure &= 2*\frac{Precision*Recall}{Precision + Recall}\nonumber \\
Accuracy &= \frac{True Negative + True Positive}{Total Number of Users}\nonumber
\end{align}

\begin{figure}[!ht]
\centering
\includegraphics[width=3in]{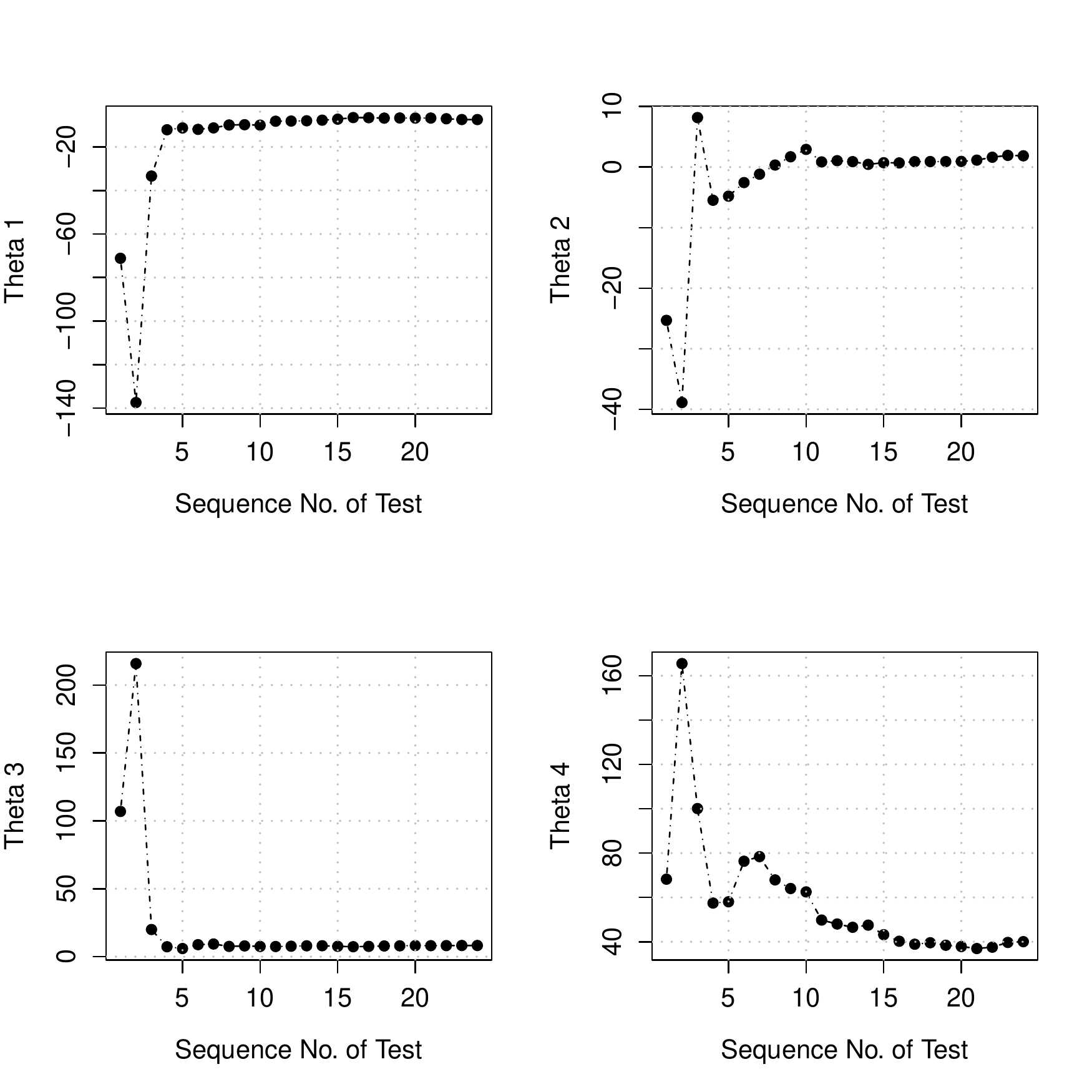}
\caption{Adaptive changes of model parameters over time}
\label{fig:changingtheta}
\end{figure}

\begin{figure}[!ht]
\centering
\includegraphics[width=3in]{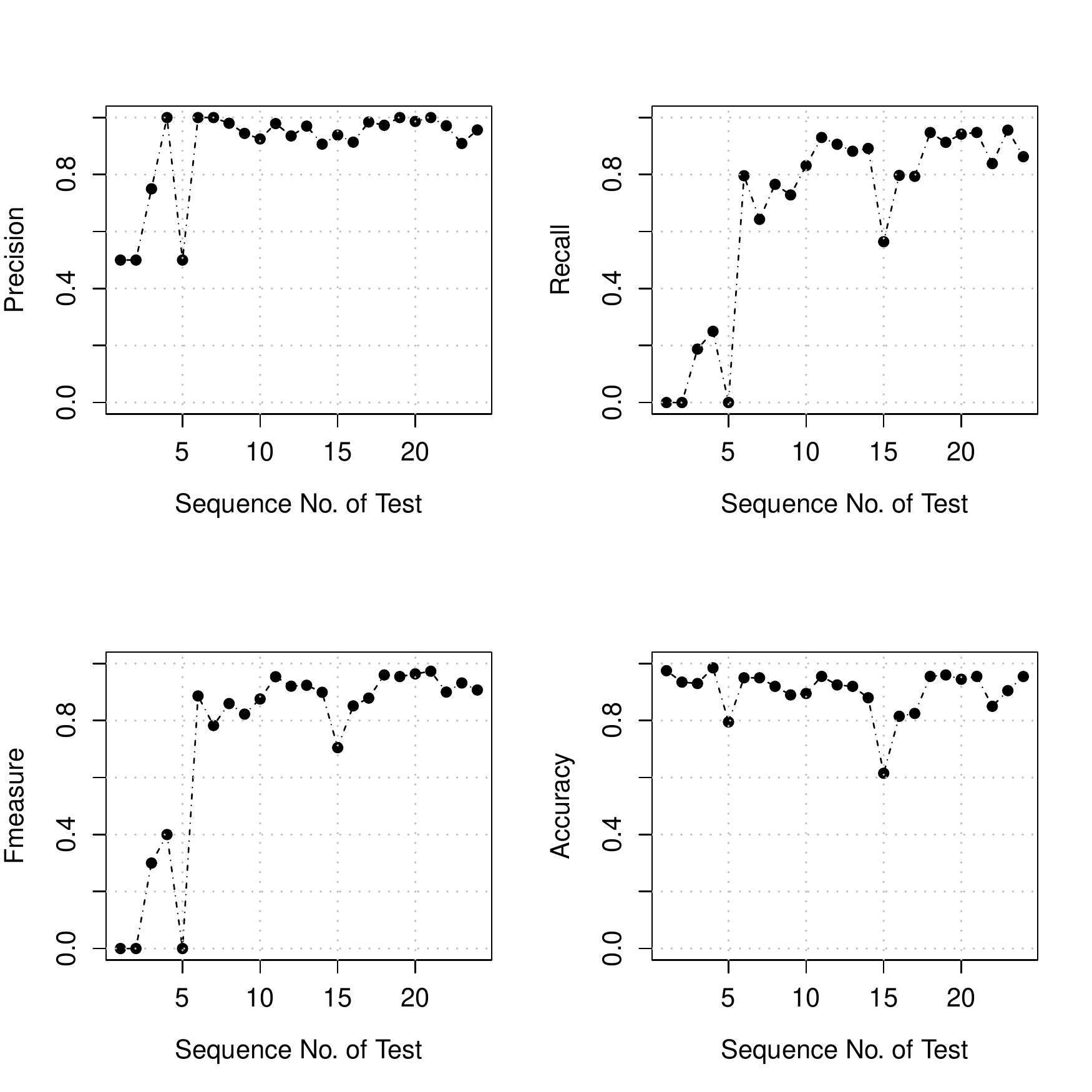}
\caption{The performance of the online detection system over time}
\label{fig:changingmetrics}
\end{figure}

Figure~\ref{fig:changingtheta} and Figure~\ref{fig:changingmetrics} show the update of model parameters and the detection performance in each iteration, respectively. In Figure~\ref{fig:changingtheta}, the four ``Theta" from $1$ to $4$ are the parameters for intercept term, SGqID, SGaID, and SGtext, respectively. We can observe that the detection model tends to converge after enough sessions have been added into the database over several iterations. For example, after 10 iterations, the precision achieves $85\%$ - $90\%$. 

We also notice that there is a ``degraded" point at the 15th iteration in the recall, f-measure and accuracy figures. After carefully checking the log file of this iteration, we find out the True/False Positive and True/False Negative of this iteration, as listed in Table~\ref{tab:15it}. We can see that the False Negative is very high, which means a large number of campaign sessions is classified as the normal ones. Nonetheless, the system is able to recover from the bad performance and works well over all measures after more Q\&A data is taken into account in training the model. The four metrics are all above $80\%$ during the last few iterations. This test scenario is similar to the practical application where we predicate the unknown sessions using current knowledge and train a new model based on the sessions after we manually label them.   

\begin{table}[!htp]
\centering
\caption{Test results of the 15th iteration}
\begin{tabular}{|c|c|c|c|}
\hline
TP & FP & TN & FN \\\hline\hline
$92\%$ & $6\%$ & $31\%$ & $71\%$ \\\hline
\end{tabular}
\label{tab:15it}
\end{table}

\begin{figure}[!ht]
\centering
\includegraphics[width=3in]{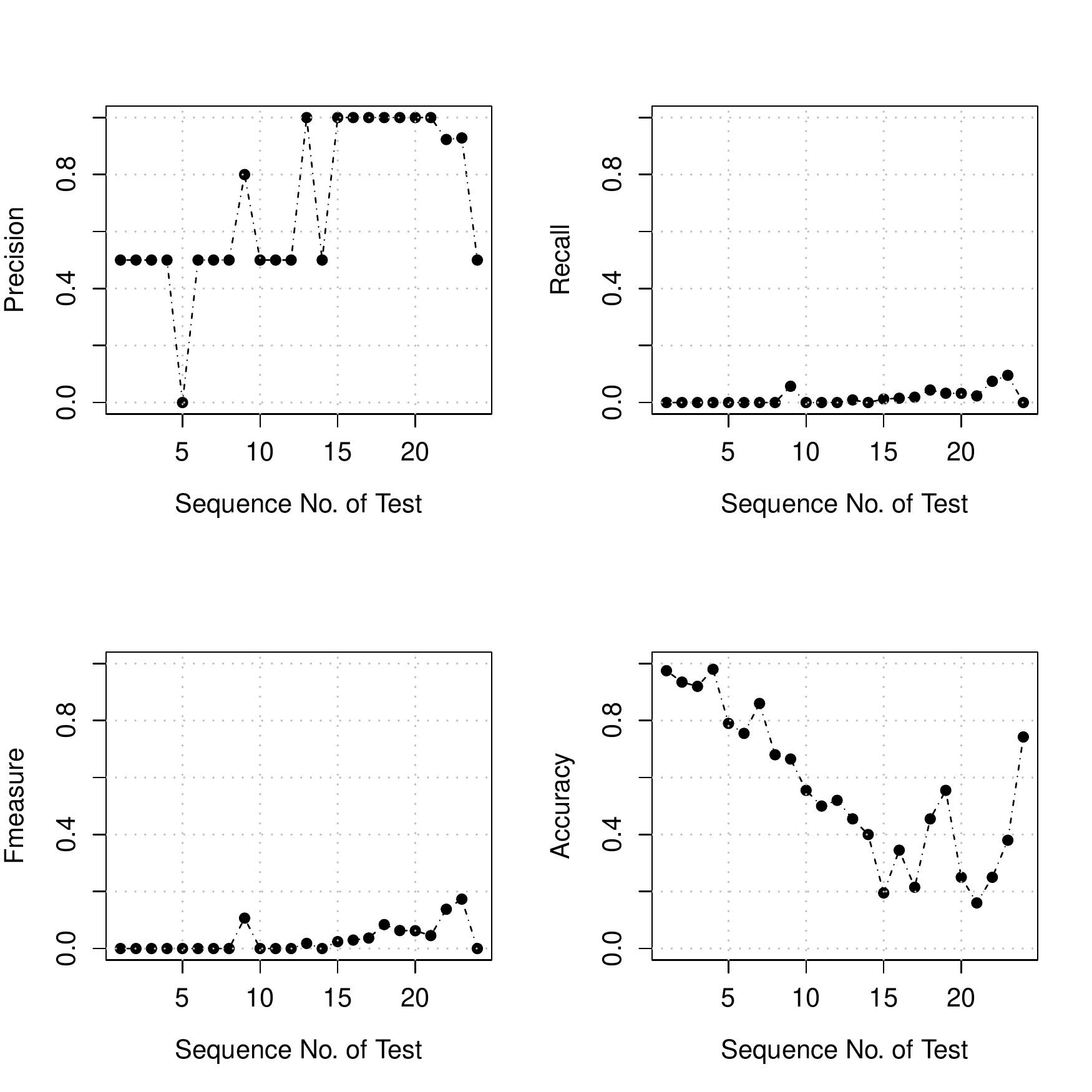}
\caption{The performance of the fixed model}
\label{fig:fixedmodel}
\end{figure}

To illustrate the advantage of adaptiveness, we also perform another test in which we fix the model after it is trained on the initial dataset. We use $200$ samples as the initial training data and build a model. We fix the model parameters, and at each iteration, we test $200$ new sessions using the fixed model. The results are shown in Figure~\ref{fig:fixedmodel}.

Since the parameters of the fixed model is only trained on the first set of training samples, we omit the figure for the system parameters. The precision in some tests is nearly 50\% and it even becomes very high in a few tests from the 15th to the 20th iterations. However, compared to Figure~\ref{fig:changingmetrics}, we note that the recall values are always very low. It means that the false negative is high. The low f-measure values confirm this problem in the fixed model. 
It means that the non-adaptive model classifies many campaign Q\&A sessions as the non-campaign sessions. Consequently, although the precision is high, other metrics indicate that the non-adaptive model has obvious bias in classification. What's worse, this model cannot update itself by new samples because the parameters are only trained on the initial training dataset. Therefore, making the predication model adaptive is critical for accurate predication in practice. 


\section{Related Work}
\label{sec:rw}

Our research is mostly related to work on spam detection and recognizing experts or authoritative users and trustworthy content in the social media. We discuss prior work on two aspects.

\subsection{Retrieving High-Quality Answers in CQA Sites}

Jeon {\em et al.}~\cite{Jeon:2006:FPQ:1148170.1148212} attempted to predict the quality of answers in a community based question answering service with only non-textual features, such as \textit{Answerer's Acceptance Ration}, \textit{Answer Length} and \textit{User's Recommendation}. They assumed the user feedback was a reliable source for the evaluation. Jurczyk {\em et al.}~\cite{Jurczyk:2007:DAQ:1321440.1321575} presented a study of link structure of Yahoo! Answers. They adopted an adaptation of the \textit{HITS} algorithm~\cite{Kleinberg:1999:ASH:324133.324140} for finding experts in the Q\&A portal. Their research was based on the assumption that the user feedback could be used to assign weights on the edges of their graph representing user relationships. 

Liu {\em et al.}~\cite{Liu:2008:USA:1599081.1599144} applied their automated summary technique to summarize answers for questions which ask for opinions. Bian {\em et al.}~\cite{Bian:2008:citeulike:2803519} tried to use both relevance between questions and answers and the quality of answers to retrieve good answers for a user query. Later, in another work by Bian {\em et al.}~\cite{Bian:2008:FBV:1451983.1451997}, they considered the effect of vote spam attacks. Such activities involved malicious voting for specific answers to improve their ranking and to decrease the ranking of competitors at the same time. Agichtein {\em et al.}~\cite{Agichtein:2008:FHC:1341531.1341557} studied the basic elements of social media and combined three features of the social media to facilitate the task of identifying high quality content, namely intrinsic content quality, interactions between users and content usage statistics. 






Fichman~\cite{Fichman:2011:journal} conducted a comparative study of answer quality on multiple Q\&A websites. Accuracy, completeness and verifiability were used as the quality measures for cross platform comparison. Fichman found that the quality of answers was significantly improved only in terms of answer completeness and verifiability, not the answer accuracy.

\subsection{Work on Crowd-Sourcing Spams in Different Realms}
Previous research has also investigated the crowd-sourcing spam in other areas. Jindal {\em et al.}~\cite{Jindal:2008:citeulike:2402521}, Ott {\em et al.}~\cite{Ott:2011:FDO:2002472.2002512} and Arjun {\em et al.}~\cite{Mukherjee:2012:SFR:2187836.2187863} attempted to detect fake review or opinion spam in the online shopping stores, like Amazon's online store. Similar to research in CQA websites, they also used textual similarity features and user-oriented features, like ratings and history records. Huang {\em et al.}~\cite{huang-yang-zhu:2011:IJCNLP-2011} developed a regression model with features suggesting quality-biased short text in Microblogging service, Twitter. They judged the quality of tweets based on relevance, informativeness, readability, and politeness of the short content and assigned different scores from 1 to 5. However, they didn't explicitly present how they define a spam-like tweet. Huang {\em et al.}~\cite{Huang:2010:DCS:1927585.1927611} conducted a similar study of commercial spam on blogging sites. They showed that the propaganda of some products in the comment of a blog post was crucial in detecting the malicious comments. The propaganda appeared in the form of URL, phone number, E-mail address, MSN numbers etc.

\nop{
} 

\nop{
}

\section{Conclusions}
\label{sec:conclusion}
Detection of hidden campaigns can improve the user's experience when using current social websites. In this paper, we disclose the behavior of a specific group of online paid posters who create commercial campaigns on the community Q\&A websites. We collect real-world datasets and identify effective features to distinguish normal sessions and the campaigns. The performance of our classifier, with integrated statistic and semantic analysis, is quite promising on the real-world case study. Based on a learning technique, we also implement a prototype of adaptive online detection system which can retrieve the result in real time. The campaign scores and/or predicated labels can help users make better decisions when searching for answers on CQA portals and help the questioners select better answers as well.

This work is our first effort to detect online paid posters of CQA websites. In the future, we will test more features to improve the adaptive performance.

\nop{
\section*{Acknowledgment}
We thank Natural Sciences and Engineering Research Council of Canada (NSERC) for partial funding support and Mathematics of Information Technology And Complex Systems (MITACS) for supporting the fourth author's internship at the University of Victoria. 
}

\nop{
One disadvantage of the study is that it relies on the exact named entity recognition. Consider that except the public crowd-sourcing websites, there are underground business assigning missions that we know nothing about. That is, how the detecting system react when new products' names appear in the Q\&A sessions?
}

\nop{
section{Issues}
On this site, the asker should register but can post as anonymous. While the answer doesn't need to have an account. In this case, we cannot use the user's history data. 

This work focuses only on the Q\&A sessions, not the users whether they are paid posters or normal users.

We will also work on the prototype and possibly make it public.

}
%
\bibliographystyle{IEEEtran}
\bibliography{Reference}

%
%

\end{document}